\documentclass[preprint,12pt]{elsarticle}

\usepackage{graphicx}

\usepackage{amssymb}

\usepackage{natbib}

\makeatletter 
\renewcommand\@biblabel[1]{}
\makeatother 

\def\lap{\;\rlap{\lower 2.5pt
\hbox{$\sim$}}\raise 1.5pt\hbox{$<$}\;}

\journal{New Astronomy}

\begin{document}

\begin{frontmatter}



\title{NBSymple, a double parallel, symplectic N-body code running on Graphic Processing Units}

\author{R. Capuzzo-Dolcetta, \cortext[cor1]{Dep. } A. Mastrobuono-Battisti, D. Maschietti}

\address{Dep. of Physics, Sapienza, Univ. of Roma, P.le A.Moro 2, I-00186, Roma, Italy}

\begin{abstract}
We present and discuss the characteristics and performances, both in term of computational speed and precision, 
of a numerical code which numerically integrates the equation of motions of $N$ `particles' interacting via Newtonian gravitation and move in an external galactic smooth field. The force evaluation on every particle is
done by mean of direct summation of the contribution of all the other system's particle, avoiding 
truncation error. The time integration is done with second-order and sixth-order symplectic schemes.  The code, NBSymple, has been parallelized twice, by mean of the Computer Unified Device Architecture to make the all-pair force evaluation as fast as possible on high-performance Graphic Processing Units NVIDIA TESLA C 1060, while the O($N$) computations are distributed on various CPUs by mean of OpenMP Application Program.
The code works both in single precision floating point arithmetics or in double precision. 
The use of single precision allows the use at best of the GPU performances but, of course, 
limits the precision of simulation in some critical situations. We find a good compromise in using 
a software reconstruction of double precision for those variables that are most critical for the overall precision
of the code.
The code is available on the web site astrowww.phys.uniroma1.it/dolcetta/nbsymple.html 
\end{abstract}

\begin{keyword}
Gravitation; Stellar dynamics; Methods: N-body simulations; Methods: numerical
\PACS 45.50.Jf 
\PACS 45.50.Pk
\PACS 95.10.Ce
\PACS 98.10.+z
\PACS 95.75.Pq

\end{keyword}

\end{frontmatter}


\section{Introduction}
\label{}




The study of systems where each particle interacts with each other is
the aim of a huge number of scientific fields (e.g. molecular dynamics,
stellar dynamics, etc.).\\
The so called classical $N$-body problem, in particular, is of great relevance
in astrophysics. Actually, the classical gravitational (or Coulombian) $N$-body problem is not solvable (in its
general form) analitically, so it is compulsory relying on numerical
simulations.
The direct all-pairs approach is formally simple but very expensive in terms of
CPU time. The CPU time grows proportionally to the number of distinct pairs in the
$N$-body system, $N(N-1)/2$, i.e., for large $N$ it scales as $N^2$, i.e.the computational complexity is $O(N^2)$.
Due to this, to reduce computational load one usually resorts on far-field
approximations or other approximation techniques for the large scale
contribution to the force on a given element (that we call \lq particle\rq ,
also in the case of a star in a stellar system), limiting direct summation to a
resticted number of neighbours.\\
In this paper we approach the study of the evolution of star clusters in
our Galaxy, as $N$-body systems where an external galactic gravitational field,
represented as an analytical expression, is summed to the interaction among all
the stars of the cluster. We evaluate the all-pairs component by direct summation,
avoiding spurious approximations; in order to speed up computations,
we exploited a double-parallelization on CPUs and on the hosted Graphic
Processing Units (GPUs), which are modern pieces of hardware capable of
very high computing speed. \\
The platform we used is made by a 2 Quad Core Intel Xeon $2.00GHz$
workstation and a GPU NVIDIA TESLA C1060.
This GPU of the last generation supports both single-precision (SP, 32-bit)
and double-precision (DP, 64-bit) floating point arithmetic. At
present, each of the processing units in the TESLA C1060 contains one DP processor
alongside the 8 SP processors; the TESLA C1060 GPU has 240 threads, meaning that
30 threads are at work when fully exploiting DP calculations.\\
Parallelization on CPUs has been done by mean of OpenMP Application Program,
supporting shared-memory parallel programming, while parallelization on
the GPUs makes use of the NVIDIA native programming language CUDA (Computer
Unified Device Architecture).\\
The OpenMP parallelization is mainly devoted to speed up\\ the time-integration, 
performed by $2^{nd}$ and $6^{th}$ order symplectic algorithms, and
the evaluation of the acceleration due to the external potential, which are both O($N$) tasks,
while CUDA was involved in the pairwise forces evaluation, whose computational
load scales as $N^2$.\\
In order to check reliability, precision and speed of our composite code,
we realized various versions of the same basic code. A comparison of these different implementations
allows us to evaluate how convenient is the usage of graphic cards coupled to
multiple CPUs for these types of simulations.\\

\section{The $N$-body problem}
In a system of $N$ gravitating point-masses, the force on the $i-th$ body of
mass $m_i$ and position vector ${\bf r}_i$ in a cartesian reference frame is the sum of the individual 
forces, ${\bf F}_{ij}$, due to all the other $j=1,2,...,N,~j \neq i$ bodies as given by

\begin{equation}
{\bf F}_{ij}=G\frac{m_i m_j}{|{\bf r}_j-{\bf r}_i|^3}({\bf r}_{j}-{\bf r}_{i}),
\end{equation}
such that the total force acting on $i$, ${\bf F}_{i}$, is

\begin{equation}\label{eq:F_i}
{\bf F}_{i}=\sum^N_{j=1, j \not= i} {\bf F}_{ij}=Gm_i\sum^N_{j=1, j \not= i}
\frac{m_j}{|{\bf r}_{ij}|^3} {\bf r}_{ij},
\end{equation}
where we set ${\bf r}_{ij}\equiv {\bf r}_{j}-{\bf r}_{i}$.
The anti-symmetry condition ${\bf F}_{ij}=-{\bf F}_{ji}$ holds,
implying the linear and angular momentum conservation laws for the isolated
case. The resulting system of equations of motion, subjected to given initial
conditions, is

\begin{eqnarray}
\label{eqfirstorder}
\left\{
\begin{array}{lll}
\ddot{ {\bf r}}_i & =  {\bf F}_{i} + {\bf F}_{ext}({\bf r}_i) \\
{\bf r}_i(0) & = {\bf r}_{i0}\\ 
\dot{ {\bf r}}_i(0) & = \dot{{\bf r}}_{i0},
\end{array}\label{eq:nbodyEv}
\right.
\end{eqnarray}

\noindent where ${\bf F}_{ext}$ accounts for an external force, expressed
by ${\bf F}_{ext}({\bf r}_i)= \nabla U_{ext}({\bf r}_i)$
if conservative ($\nabla$ is the usual gradient operator acting on the external force potential,
$U_{ext}$).\\
It is well known that the existence of just 10 constants of motion,
independently of $N$, does not allow a general analytical solution for
$N\geq 3$.
This obliges to rely on numerical methods, whose application is made ackward by
the contemporary presence of problems on both the {\it small} length scales
(close
gravitational encounters) and the {\it large} ones (gravity has no screen, so
the contribution of every component
of the system, even if distant, must be included for a correct force
evaluation). This \lq double ~divergence\rq~ of the classic newtonian interaction potential has
two different consequences:
i) close encounters correspond to an unbound force between colliding bodies
($|{\bf F}_{ji}| \rightarrow \infty$ when $|{\bf r}_{ij}| \rightarrow 0$)
yielding to an
unbound error in the relative acceleration;
ii) the need of the complete force summation over the whole set of distinct
$N(N-1)/2$ pairs in the system implies an overwhelming CPU charge, practically
unaffordable even with most modern, quick CPUs.

Problem i) is often cured by mean of the introduction of a \lq softening\rq~
parameter, $\epsilon$, in the interaction potential, which assumes the smoothed
form

\begin{equation}
U_{ij}=G\frac{m_j}{\sqrt{|{\bf r}_{ij}|^2+\epsilon^2}}.
\label{softpot}
\end{equation}

The corresponding total force acting on the $i-th$ particle is

\begin{equation}
{\bf F}_{i}\simeq Gm_i\sum^N_{j=1} \frac{m_j{\bf r}_{ij}}{(|{\bf
r}_{ij}|^2+\epsilon^2)^{3/2}}.
\label{eq:F_iSoftened}
\end{equation}

In the latter sum the condition $j\not=i$ is no longer needed, because ${\bf
F}_{ii}=0$ if $\epsilon \neq 0$. Note that the introduction of the softening
parameter corresponds to substitute point masses with Plummer's spheres (Plummer 1911),
where the mass $m_i$ is distributed around the centre according to the
density law

\begin{equation}
\rho_i(r)=\frac{3m_i}{4\pi \epsilon^3}\frac{1}{(\epsilon^2+
r^2)^{5/2}}.
\end{equation}
The reader interested to a deeper discussion of the pair interaction
mollification may refer to Sect. 2 of the Capuzzo--Dolcetta et al. (2001) paper.
It is relevant recalling that, in any case, the chance of close encounters is
larger for small values of $N$, when the ``granularity" of the
system is more significant.

Throughout this paper we will use the softened interaction given by Eq. 5 with $\epsilon=0.005(R/N^{1/3})$, where
$R/N^{1/3}$ is a measure of the average distance of a particle of the $N$-body system to its closest neighbour.

Problem ii) is faced in many different ways; we do not want to cite here them all
in order not to be too long. We just say that the common line is introducing
averaging (mean-field) methods, dividing the force acting on a particle
and due to the rest of the system into a ``large" scale, slowly varying,
coarse-grain, contribution ($ {\bf F}_{ls}$) and into a ``small" scale, rapidly
varying, fine-grain, contribution, which is represented as a summation limited to a set
of $n<N$ neighbouring particles, to give

\begin{equation}\label{eq:F_iSoftening}
{\bf F}_{i}\simeq Gm_i\sum^n_{j=1} \frac{m_j{\bf r}_{ij}}
{(|{\bf r}_{ij}|^2+\epsilon^2)^{3/2}} + {\bf F}_{ls} ({\bf r}_i).
\end{equation}

Of course, setting $n=N$ (which means that the neighbouring sphere contains all
the system particles) makes the approximated expression above equal to the correct direct 
summation; in this case ${\bf F}_{ls}$ is contributed by an external force, only.

\section{The $N$-body code}
\label{sect3}
The main scientific aim of this paper is producing a high precision, fast and reliable code to study the evolution of stellar systems of the size of open clusters and globular clusters.
Our goal is the study of their dynamical evolution taking into account both the internal mutual force
and the effect of the external galactic field.

The code generates, first, the initial conditions for the $N$ particles of the system,
whose individual masses are chosen by a given mass spectrum. For the scope of
this paper, which aims, mainly, at investigating the code quality and performances in different 
hardware and software environments, we gave, for the sake of simplicity, all the particles 
the same mass $m_i=m$ $(i=1,2,..,N)$. The total mass, $M=Nm$, is assumed as mass unit.
For the same purpose of simplicity we give particles an initial spatially
uniform distribution within a sphere of given (unitary) radius, $R$, with
velocities, also, uniformly distributed in direction and absolute values
and rescaled, in their magnitude, to reproduce a given value of the virial ratio
(we remind that the virial ratio is defined as $Q=2K/|\Omega|$, where $K$ and
$\Omega$ are, respectively, the system kinetic and potential energies;
for a stationary system, $Q=1$). Note that the further assumption $G=1$ in the
equations of motion implies that the ~\lq crossing\rq~ time
$T=(GM)^{-1/2}R^{3/2}$ is the unit of time.

The forces due to the mutual interaction among stars in the cluster
(internal forces) are  given by Eq. 5 and summed 
to the effect of the external galactic field.
The force due to the external field (${\bf F}_{ext}$, second term in
Eq. 3) is accounted by an analytical expression for the Galactic
potential as given by Allen and Santill{\' a}n (1991). In this paper the authors consider the
Galactic potential as given by three components: a bulge, a disk, and a halo.
The bulge and the halo have a spherical symmetry, while the disk is
axisymmetric.

\subsection{Time integration}

It is well known that ordinary numerical methods for integrating Newtonian
equations of motions become dissipative and exhibit incorrect long term
behaviour. This is a serious problem when facing $N$-body problems, particularly when 
studying them on a huge time baseline. One possibility is using symplectic integrators.\\
Symplectic integrators are numerical integration schemes for Hamiltonian
systems which conserve the symplectic two-form ${\rm d}{\bf p}\land
{\rm d} {\bf q}$ exactly, so that
(${\bf q}(0),~{\bf p}(0))$
$\to({\bf q}(\tau),~{\bf p}(\tau))$ is a canonical transformation. They are
characterized by
time reversibility.
For both explicit and implicit integrators, it was shown (Yoshida 1991) that the
discrete
mapping obtained
describes the exact time evolution of a slightly perturbed Hamiltonian system
and thus possesses
the perturbed Hamiltonian as a conserved quantity. This guarantees that there is
no secular change
in the error of the total energy (which should be conserved exactly in the
original flow) caused by
the local truncation error. If the integrator is not symplectic, the error of
the total energy grows
secularly, in general.\\
Our code allows the choice of two different symplectic methods.
One is the simple, classic, ``leapfrog" method, which is $2^{nd}$-order
accurate; the other is a more accurate $6^{th}$-order explicit scheme whose
coefficients are taken from the first column of the Table 1 (SI6A) of  Kinoshita,Yoshida, \& Nakai (1991), which leads to a conservation of energy for a factor fifty better than that with the
other two possible sets of coefficients. 
Of course, the $6^{th}$-order symplectic integrator is much slower than the leapfrog, 
requiring 7 evaluations of force functions per time step, like, for instance, in a $6^{th}$-order
Runge Kutta method.\\

\begin{figure}
\begin{center}
\includegraphics[width=\textwidth]{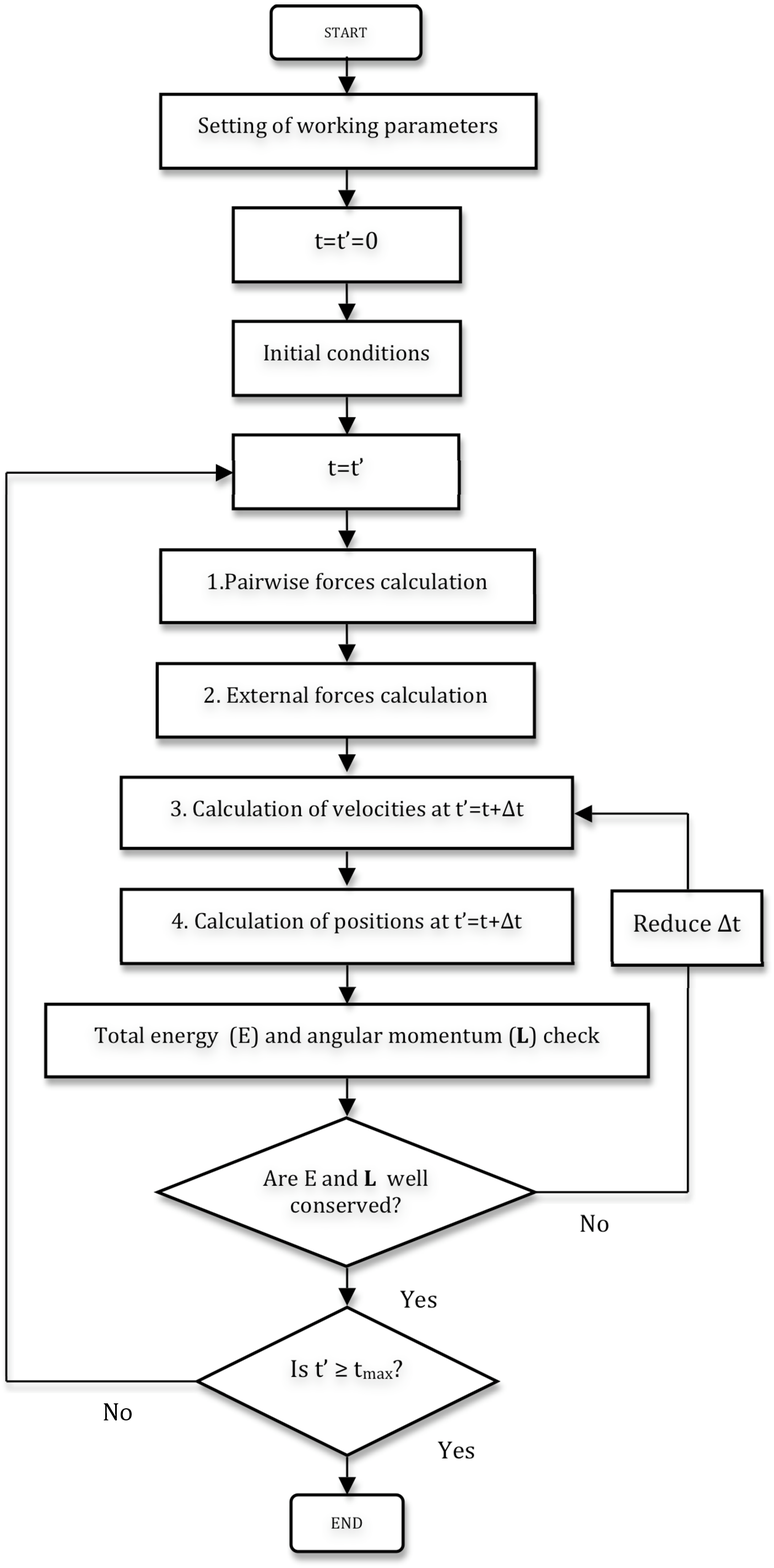}
\caption{Flowchart of our basic code}
\label{fig1}
\end{center}
\end{figure}

\section{Implementation of the code}

\subsection{Hardware and software}
\label{hw&sw}


As we said before, the direct evaluation of the forces among particles in an $N$-body 
system is quite a heavy task for computers, mainly because of the computation 
of the elements of the $|{\bf r}_{ij}|$ (euclidean distance) array,
whose number of distinct elements is $N(N-1)/2$ (due to the symmetry $|{\bf r}_{ij}|=|{\bf r}_{ji}|$) 
thus scalings as $N^2$. As a matter of fact, every distance computation requires 
the evaluation of a square root, which is an expensive computational task.
Actually, every pair force vector, ${\bf F}_{ij}$,
requires $20\div 30$ floating point operations. This means that time integrations
of $N$-body systems, when extended over a time interval large enough to have a scientific relevance, 
overwhelm the power of every single CPU platform for $N$ above few thousands. If one wants to keep direct summation, without relying on approximations like, for instance, mean-field techniques, the only possibility
is resorting on parallel computing and/or dedicated machines.\\
It is out of the purposes of this paper the discussion of parallelization of $N$-body codes on large main 
frames, which is not a trivial task, for the slow decay with distance of the gravitational interaction. 
Here we deal with the problem of implementing an efficient $N$-body integrator on our computing CPU+GPU platform.\\
The particular workstation we used has two Quad Core Intel Xeon processors,
each running at $2.00$ GHz, $4$GB DDR2 RAM at $667$ MHz and two NVIDIA TESLA
C1060 GPUs, connected to the host via two slots PCI-E 16x.
\\
NVIDIA TESLA C1060 has 240 processors, each of them has a clock of 1.296 GHz, for a nominal
SP floating point peak performance of 933 GFlops/s (78 GFlops/s in DP).\\
Some authors have already faced the problem of implementing gravitational $N$-body integrations 
on GPUs with different approaches (Barsdell et al. 2009). 
Here we explain how we get our own implementation, 
leading to a code, which we call NBSymple (ackronym for NBody Symplectic integrator).
Various versions of the same basic code have been realized.\\
The first is fully serial (NBSympleA), i.e. it runs on a single
processor. This code constitutes a \lq unit of measure\rq~ of the
performances of the various parallel versions.\\
In a second version of the code (NBSympleB) we implemented parallelization,
with Open Multi-Processing (OpenMP) directives, of both the $O(N^2)$ pairwise
interactions and the $O(N)$ calculations (i.e. the time integration and
evaluation of the Galactic component of the force on the system stars)
over the double Quadcore host.
OpenMP is an application programming interface (API) that supports
multi-platform shared memory
multiprocessing programming in C, C++ and Fortran
on many architectures, including Unix, Linux and Microsoft Windows platforms. It
consists of a set of compiler directives, library routines, and environment
variables that influence run-time behavior.\\
In the third  version (NBSympleC) the all-pairs interactions ($O(N^2)$ load
calculations) are demanded to the NVIDIA TESLA C1060 GPU, using Compute
Unified Device Architecture (CUDA, see\\
{\it http://www.nvidia.com/object/cuda\_home.html})
while all the remaining tasks are done by a single (or multiple) CPU. CUDA is a parallel
computing architecture developed by NVIDIA that extends C by allowing the programmer to define C
functions, called kernels, that, when called, are executed $n$ times in parallel by $n$ different CUDA
threads. CUDA programming language is basically a C programming language extended with a number
of keywords.
CUDA threads may execute  on a physically separate device that operates as a
coprocessor to the host running the C  program. In this case, the rest of the
code runs on a single CPU.\\
In another (fourth) version of the program (NBSympleD) we again use CUDA to
evaluate the $O(N^2)$
portion of the code (like in NBSympleC), while the $O(N)$ computations (time
integration and evaluation of the acceleration due to the Galaxy) are parallelized sharing work between all
the eight cores of the host, using OpenMP in the same way done in NBSympleB.\\
The last (fifth) implementation (NBSympleE) uses CUDA on one or two GPUs to
evaluate the total force over the system stars, i.e. both the all-pairs component
and that due to the Galaxy. In NBSympleE, the O(N) computations were shared into an OMP part (time integration) and into a GPU part (smooth galactic force contribution) to maximize the efficiency. 
\\
Table 1 summarizes the above (main) characteristics of the various versions of 
the code.

\begin{table}[!h]
\small
\begin{center}
\begin{tabular}{lccc}
\hline
{\bf Code Version}            &    {\bf  1PE}        &    {\bf  OMP (8PEs)} & {\bf GPU}\\
\hline
NBSympleA                          &     $N+N^2$        &        -          &    -      \\
NBSympleB                         &        -        &$N+N^2$          &    -       \\
NBSympleC                        &       $N$            &        -          &$N^2$       \\
NBSympleD                         &        -        &$N$              &$N^2$    \\
NBSympleE                        &        -        &$N$               &$N+N^2$\\
\hline
\end{tabular}
\end{center}
\caption{Synoptic table summarizing the computational complexity of the tasks performed by the various 
processing units as exploited by the different versions of our NBSymple code (see Sect. 4.1).}
\label{tab1}
\end{table}

\subsection{The code structure}
The $N$-body integration scheme consists of two main parts.
In the first, given positions and velocities of all the $N$ bodies, the forces
between stars and that due to the overall, smooth, Galaxy distribution are evaluated. Consequently, in
the second part, the code predicts the velocities and positions of the particles by mean of the
previously calculated accelerations (see flow-chart in Fig. 1).\\
In all of the five implementations of the code the advancing in time of
velocities and positions of stars of the system is performed by the CPUs.
This choice is motivated by that time advancing is more sensitive to
round-off errors and the time needed for its computation grows only linearly with $N$, making
convenient using the double precision representation available on the CPUs. Actually, TESLA C1060 
supports double-precision floating point arithmetics, but in this case performances decay significantly
with respect to using single precision. Another problem is due to the bandwidth available for the data exchange between GPUs and CPUs.
Performing time integration on GPUs would require sending and receiving a huge
amount of data from GPUs to CPUs, with a significant decay of performance.\\
We now give a short description of how the code implementations that use CUDA 
handle communication of data between the host (CPU) and the device (GPU).\\
Before we invoke the kernel we copy the positions of the $N$ particles on the GPU's global memory; after 
the calling we take the accelerations, calculated on the GPU device, and send them to the CPU
memory. The kernel and device functions used to evaluate the forces are very
similar to those described in Nyland et al. (2007).
In this latter work the authors introduced the notion of a \lq computation tile\rq, as
a squared $p\times p$ sub-array of the $|{\bf F}_{ij}|$, $N\times N$ force array.
To calculate $p^2$ interactions we need the knowledge of the positions of
$2p$ bodies (for details see Nyland et al. 2007).
These \lq body descriptions\rq~ are stored in the shared memory, which has little
reading latency (4 clock cycles) respect to global memory (400-600 clock
cycles).
To achieve optimal reuse of data, the computation of a tile is arranged in a
way that the interactions in each row are evaluated in sequential order
(updating the acceleration vector),  whilst the various  rows are evaluated in parallel.\\
The interaction between a pair of bodies is implemented as  an entirely serial
computation. A tile is evaluated by $p$ threads performing the same sequence of
operations on different data. Each thread updates the acceleration on one body
as a result of its interaction with $p$ other bodies. They load $p$ body descriptions from the GPU
device memory into the shared memory provided to each thread block in the CUDA model.
Each thread in the block evaluates $p$ successive interactions. The results of the tile
calculation are the $p$ updated accelerations.\\
A thread block is defined as a collection of $p$ threads that execute some
number of tiles in sequence. In a thread block, there are $N/p$ tiles, with $p$
threads computing the forces on $p$ bodies (one thread per body). Each thread
computes all the $N$ interactions for one body. \\
The kernel is invoked on a grid of thread blocks to compute the acceleration of
all the $N$ bodies. Because there are $p$ threads per block and one thread per body, 
the number of thread blocks needed to complete all $N$ bodies is $N/p$, so we have 
a $1D$ grid of size $N/p$. The result is a total of $N$ threads that perform $N$ force 
calculations each, for a total of $N^2$ interactions. \\
The number of calculations performed by the device is $N^2$.
This is redundant respect to the actual number of distinct pairs in the system,
$N(N-1)/2$; by the way, the limitation to $N(N-1)/2$ of the number of force evaluations
requires a heavier load of internal communication and synchronizations (Belleman et al.
2008), with a resulting net performance decrease. \\
The dimension of the tile and the number of thread blocks depend on the number
of particles involved in the simulation. There are 30 multiprocessors on the
NVIDIA TESLA C1060, so the block dimension ($p$) must be such that $N/p$ is 30
or larger to use all the multiprocessors available.

The entire part of the program that performs the time integration of the system is
enclosed in a parallel section, through OpenMP directives.  We shared work
between all the CPUs available declaring variables in an appropriate way, and
using properly the OpenMP directives to parallelize the various \lq for-cycles \rq~ 
that perform the time integration and the computation of the galactic
contribution to the accelerations (both scaling as $N$). The copy of the data 
from and to the GPU and the kernel are invoked by the master thread.\\

As we said above, our GPUs allow us the use of DP floating point
representation. To reduce the bandwidth needed to transfer data between CPUs and
GPUs we constructed C structures of four DP variables, emulating
CUDA's float4. As we  see in the next Section the performances of the program
decrease a lot. On the other side, using the $6^{th}$order integrator we reach a
precision 7 or 8 orders of magnitude better than that reached with  single-precision
arithmetics.\\
Finally, we realized another version of our program to work on $n$ GPUs of
the same type, besides sharing the job on the 8 CPUs of the workstation.\\
Through OpenMP directives we created $n$ threads. Each thread communicates
with one of the device and copies the positions of all the particles on
its global memory. Moreover each of the $n$ threads, launches a kernel
on one of the $n$ available devices. So we have $n$ kernels, each of which calculates
the accelerations for $N/n$ particles (in our case $n=2$).
After this calculation, the instanced thread copies the results
on the host and the number of threads is reset to 8 to continue the time
integration.\\

\section{Results}
\subsection{Performances and accuracy of the codes}
We tested the performances in terms of both precision and
speed of our code in its various versions.
As outlined in Sect. 3, the initial conditions for the
$N$-body system are picked from a uniform phase-space distribution of
equal mass particles, with an initial virial ratio $Q=0.3$.
The particle-particle interaction potential is softened (see Eq. 4 and 5) and the system moves in the Galaxy, represented with the
Allen and Santillan (1991) model, on a quasi circular, planar orbit at 
about the same galactocentric distance of the Sun.\\
To evaluate the quality of integration we checked the time behaviour of $4$ quantities: the
total energy ($E$) and the three components of the angular momentum (${\bf L}$) vector.
The energy is a true constant of motion, while ${\bf L}$ should vary in time due to 
the external field torque, 
$\dot {\bf L}={\bf M}_{ext}$,
being a constant only in case of either isolated system or system embedded in an external 
spherically symmetric (respect to the system barycentre) potential.
In any case, the external torque is expected to be little due to the small 
cluster size, as we checked by the computation of the total torque via direct 
summation,
\begin{equation}
{\bf M}_{ext}= \sum _{i=1}^N {\bf r}_i \times {\bf F}_{ext}({\bf r}_i) 
\end{equation}
The energy and angular momentum conservation is evaluated through 
computation of 
\begin{equation}
\frac{\Delta E}{|E(0)|}=\frac{E(t)-E(0)}{|E(0)|},
\end{equation}
and
\begin{equation}
\frac{\Delta {\bf L}}{|{\bf L}(0)|}=\frac{{\bf L}(t)-{\bf L}(0)}{|{\bf L}(0)|}.
\end{equation}

Of course the quality of both energy and angular momentum conservation depends on the
code version, the best conservation performances being achieved with the use of double-precision
and $6^{th}$ order symplectic integration. Table 2 gives the run with time of the average errors per step in the case of the symplectic $6^{th}$ order time integration with DP arithmetics.

Fractional energy conservation per cycle is in the range 
[$1.4\times 10^{-14},8.8 \times 10^{-14}$],
while the $x$ and $y$ components of the angular momentum 
are conserved in the range
[$1.4\times 10^{-13},6\times 10^{-11}$].

\begin{table}[!h]
\small
\begin{center}
\begin{tabular}{cllll}
\hline
$t$ & $\langle |\frac{\Delta E}{E_0}|\rangle$&$\langle|\frac{\Delta L_x}{L_0}|\rangle$ &
$|\langle\frac{|\Delta L_y|}{|L_0|}|\rangle$ &
$|\langle\frac{|\Delta L_z|}{|L_0|}|\rangle$\\
\hline
1.25& 8.83$\times10^{-14}$ & 3.20$\times10^{-13}$ &    1.88$\times10^{-13}$ &    3.54$\times10^{-17}$\\
5& 1.44$\times10^{-14}$& 8.26$\times10^{-12}$ &    1.43$\times10^{-12}$ &    1.13$\times10^{-17}$\\
10& 1.37$\times10^{-14}$& 1.39$\times10^{-11}$ &    3.12$\times10^{-12}$ &    4.53$\times10^{-17}$\\
15& 6.87$\times10^{-14}$& 3.42$\times10^{-12}$ &    4.75$\times10^{-11}$ &    0.00$\times10^{}$\\
20& 3.45$\times10^{-14}$& 3.90$\times10^{-12}$ &    1.76$\times10^{-10}$ &    1.41$\times10^{-17}$\\
25& 8.42$\times10^{-14}$&9.32$\times10^{-11}$ &    2.42$\times10^{-10}$ &     3.68$\times10^{-17}$\\
\hline
\end{tabular}
\end{center}

\caption{Percentage relative variations in energy and angular momentum per time step in the time interval [0,25], which corresponds to about 0.4 orbital revolution around the galactic center. 
The symplectic $6^{th}$ order method is used in double-precision mode.
The errors are obtained averaging the error values in 6 time intervals composed 
by $2500$ time steps each .}
\label{tab2}
\end{table}

\begin{figure}[htbp]
\begin{center}
\includegraphics[width=\textwidth]{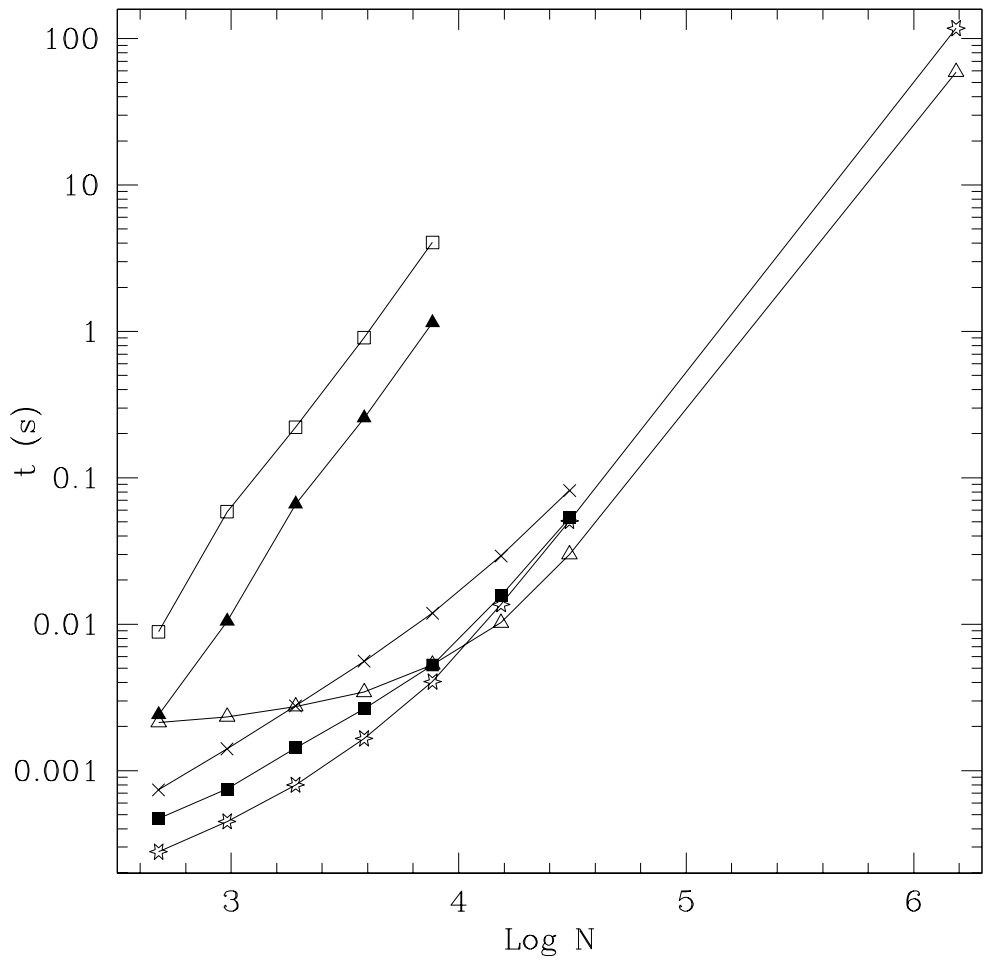}
\caption{The (averaged over $1000$ cycles) solar time (in seconds) spent
for one leap-frog integration step in single precision mode,
as a function of $N$. Line with empty squares: NBSympleA code.
Line with filled triangles: NBSympleB. Line with crosses: NBSympleC.
Line with filled squares: NBSympleD. Line with stars:
NBSympleE with a single GPU. Line with empty triangles: NBSympleE with
two GPUs.}
\label{fig2}
\end{center}
\end{figure}

\begin{figure}[htbp]
\begin{center}
\includegraphics[width=\textwidth]{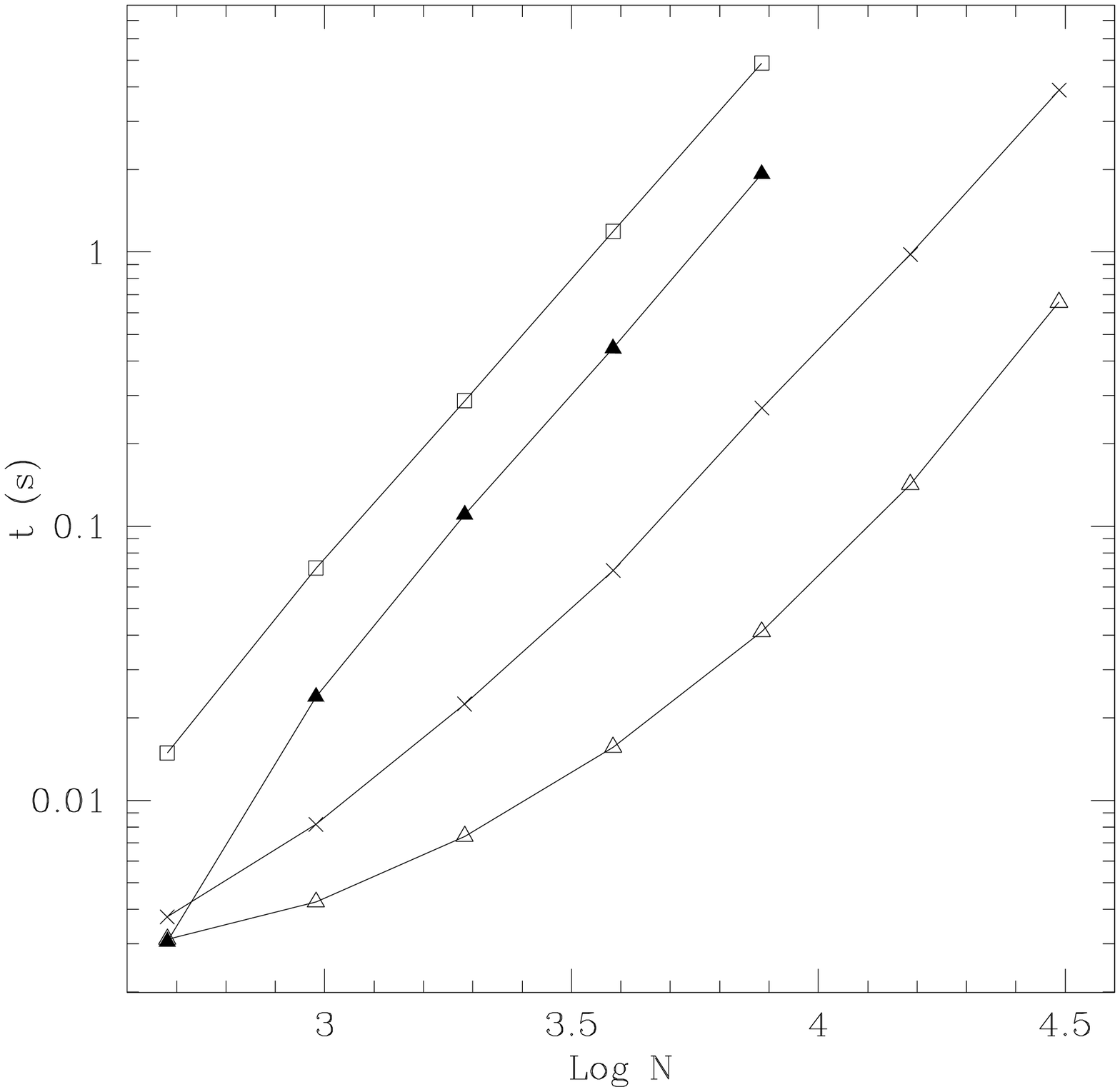}
\caption{As in Fig. 2, but for the double-precision codes.
The line with filled squares is not reported because it is
almost identical to that with crosses.}
\label{fig3}
\end{center}
\end{figure}

\begin{figure}[htbp]
\begin{center}
\includegraphics[width=\textwidth]{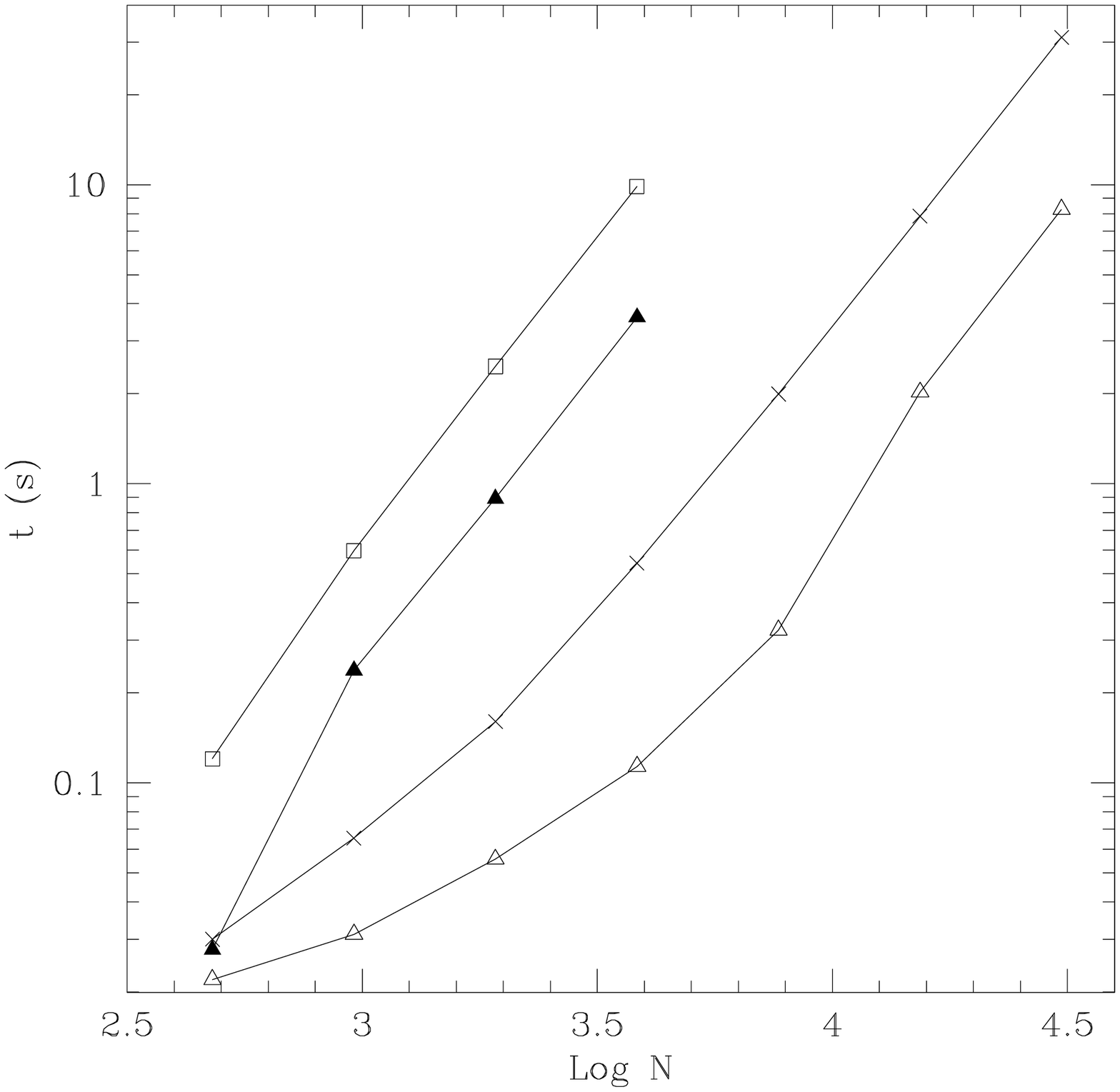}
\caption{As in Fig. 2, but for the sixth-order, double-precision
codes.
The line with filled squares is not reported because it is
almost identical to that with crosses.}
\label{fig4}
\end{center}
\end{figure}

\begin{figure}[htbp]
\begin{center}
\includegraphics[width=\textwidth]{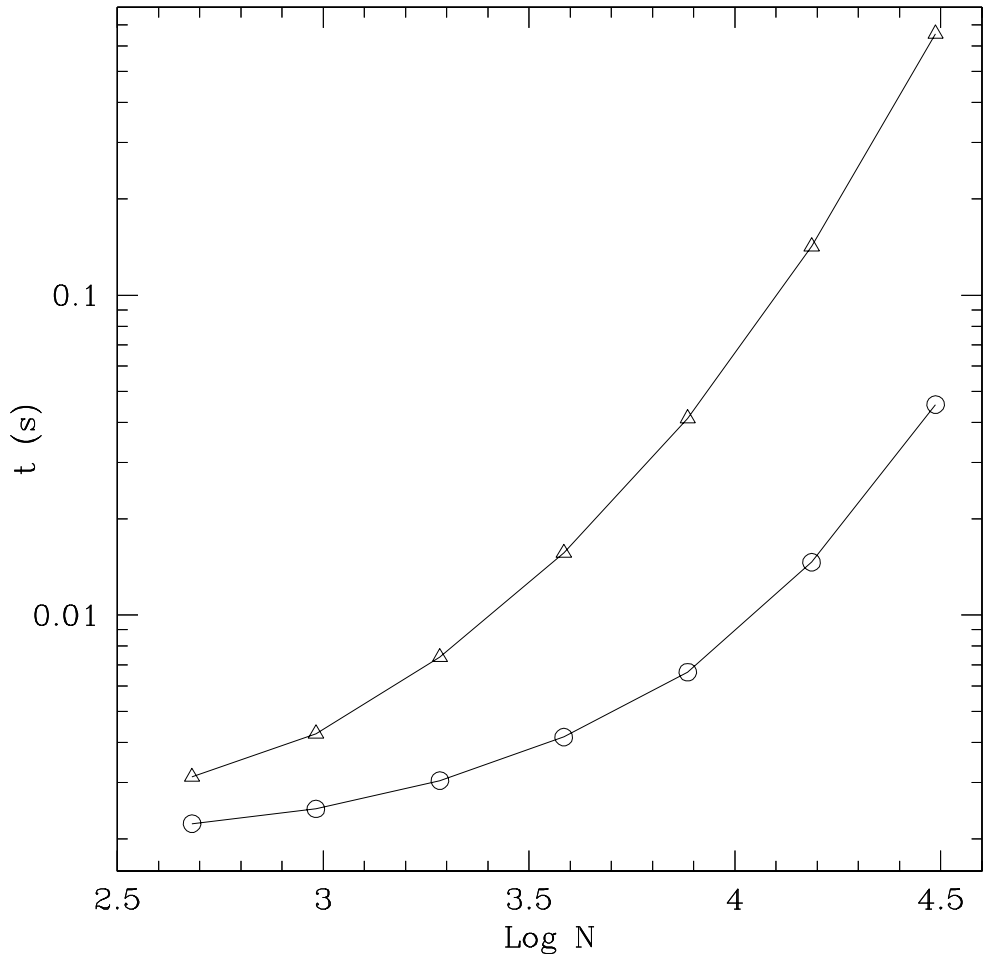}
\caption{The (averaged over $1000$ cycles) solar time (in seconds) spent for a
single integration step as a function of $N$ by the NBSympleE codes in
the leapfrog, hardware double precision mode (line with triangles) and
in the leapfrog, software double precision mode (line with circles).}
\label{fig5}
\end{center}
\end{figure}

\begin{figure}[htbp]
\begin{center}
\includegraphics[width=\textwidth]{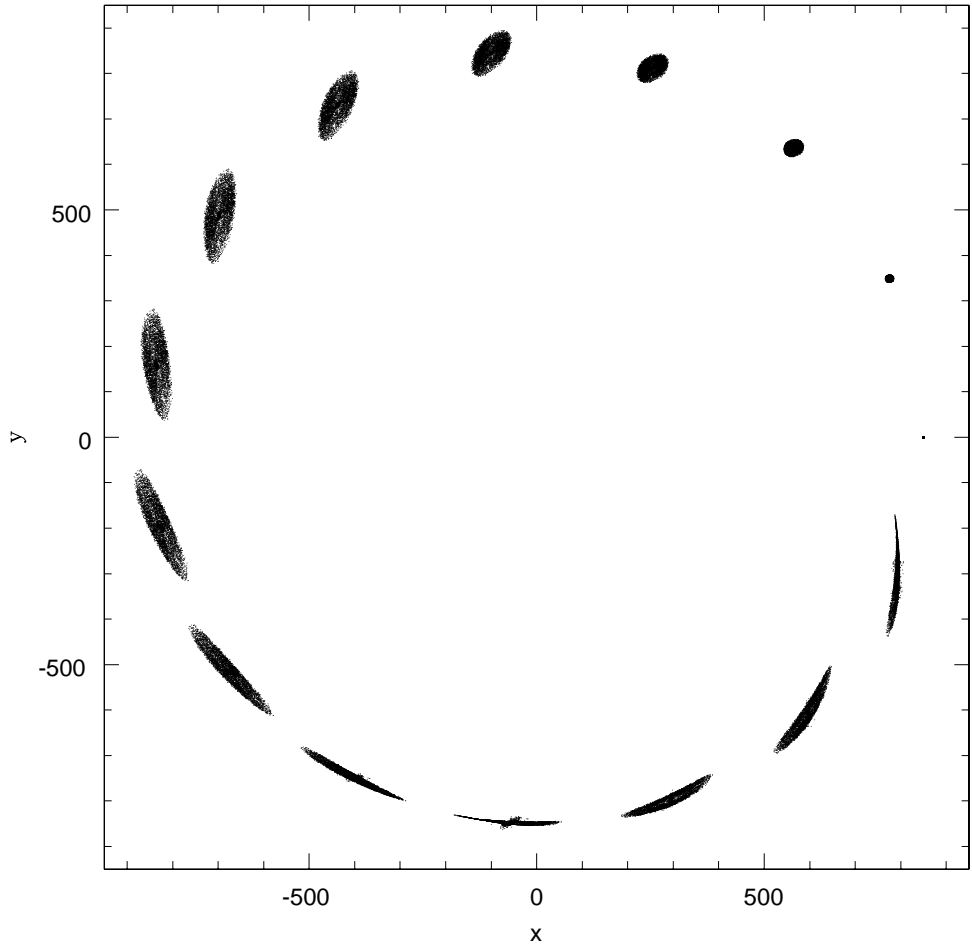}
\caption{Configurations of the $N=15,360$ simulated cluster moving on the Milky Way simmetry plane, at various times,
from t=0 to t=60. The cluster motion is counterclockwise.}
\label{fig6}
\end{center}
\end{figure}

\begin{figure}[htbp]
\begin{center}
\includegraphics[width=\textwidth]{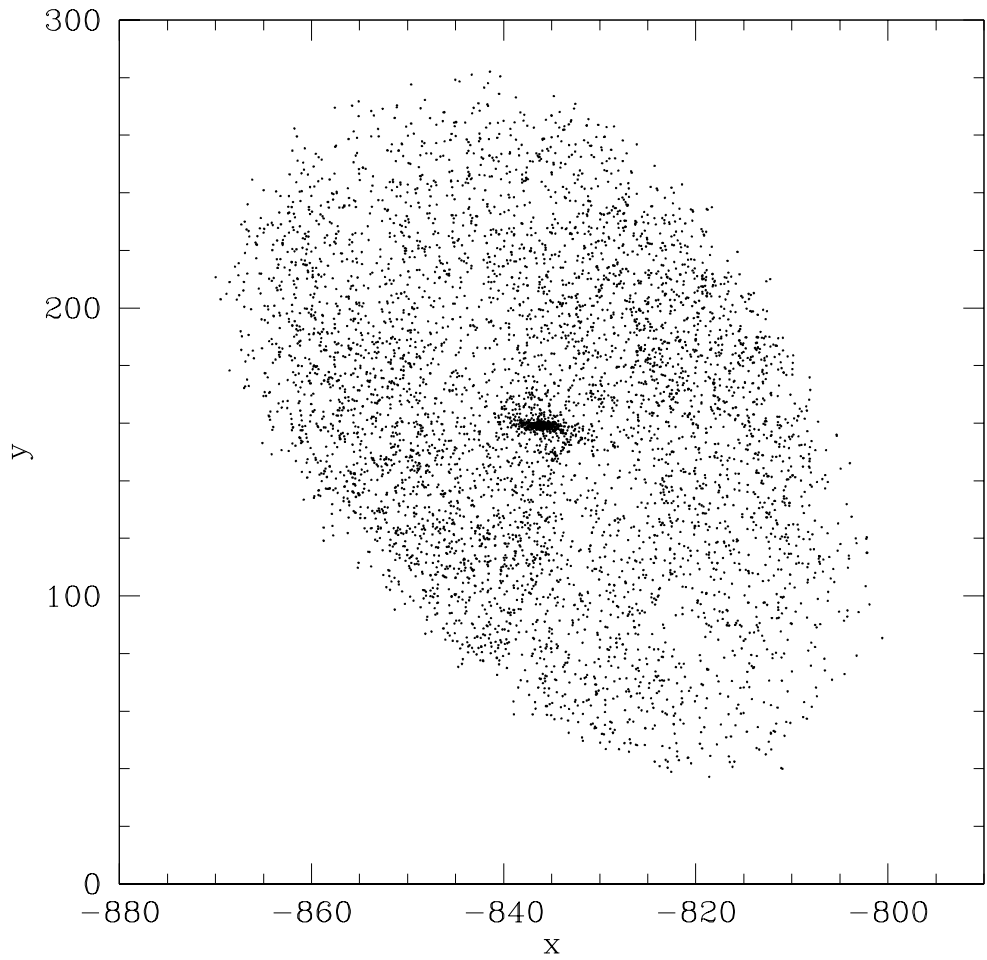}
\caption{Snapshot of the cluster on quasi-circular orbit at the time $t=29.4$ (i.e. at about
half of its first revolution around the galactic centre).}
\label{fig7}
\end{center}
\end{figure}

\begin{figure}[htbp]
\begin{center}
\includegraphics[width=\textwidth]{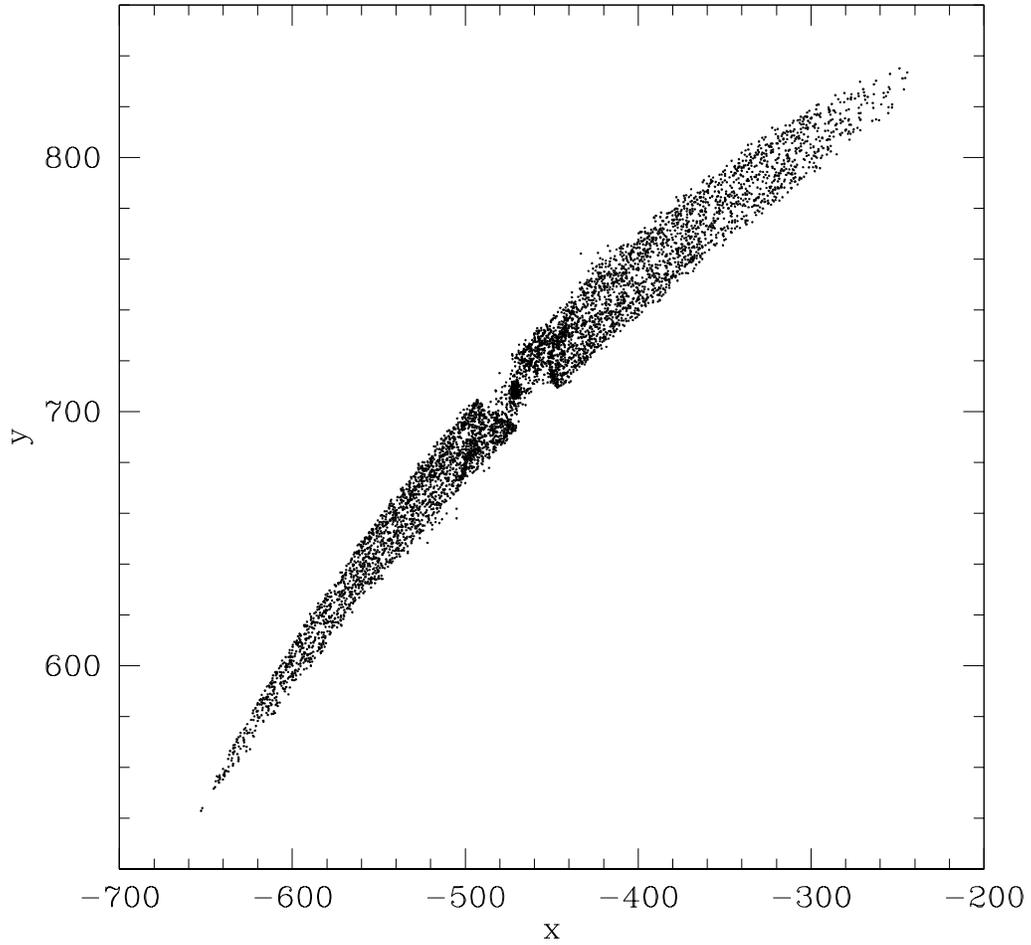}
\caption{Snapshot of the cluster on quasi-circular orbit at $t=84$ (i.e. at about $1.4$ revolutions around the galactic centre).}
\label{fig8}
\end{center}
\end{figure}

\begin{figure}[htbp]
\begin{center}
\includegraphics[width=\textwidth]{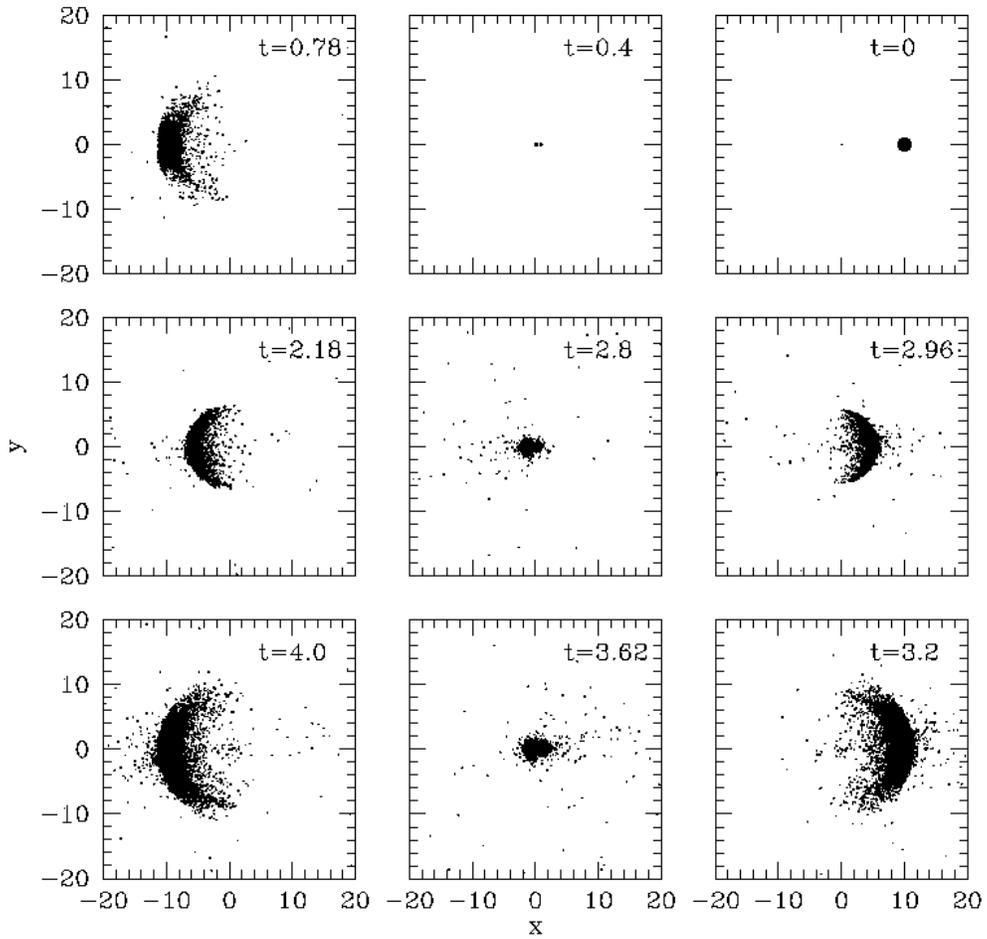}
\caption{Snapshots of the $N=15,360$ cluster plunging on quasi-radial orbit through the massive 
central object ($M_{bh}/M=10$) at the times labeled in the various panels.}
\label{fig9}
\end{center}
\end{figure}


\begin{table}[htdp]
\begin{center}
\begin{tabular}{cccc}
\hline
N                    &	480  		&	30720	& 1536000\\
\hline					
NBSympleC	&   	5.97		&	409		&	423	\\     
''			&	3.36		&	21.9		&	22.0	\\     
NBSympleE	&    	11.29		&    	797		&	846	\\     
''		 	&   	3.37		&    	43.8		&	44.0	\\     
\hline
\end{tabular}
\end{center}
\caption{Performances in GFLOPs/s of two versions of the NBSymple code (C and E) in their SP (first and third row) and DP (second and fourth row) modes. Note that NBSympleE exploits the power of two TESLA C1060 GPUs.}
\label{tab3}
\end{table}

\begin{table}[t]
\small
\begin{center}
\begin{tabular}{|c|ccc|ccc|ccc|}
\cline{2-10} \multicolumn{1}{c|}{} & \multicolumn{3}{c|}{480} & \multicolumn{3}{c|}{30,720 } & \multicolumn{3}{c|}{1,536,000 }\tabularnewline
\hline task  & C  & D  & E  & C  & D  & E  & C  & D  & E\tabularnewline
\hline 1  & 23.5  & 36.4  & 81.4  & 58.4  & 89.6  & 94.2  & 98.56  & 99.74  & 99.87 \tabularnewline
2  & 70.3  & 36.0  & -  & 38.0  & 9.0  & -  & 1.31  & 0.18  & - \tabularnewline
3  & 4.5  & 16.8  & 13.8  & 2.7  & 0.9  & 4.4  & 0.10  & 0.06  & 0.10 \tabularnewline
4  & 1.7  & 10.8  & 4.8  & 0.9  & 0.5  & 1.4  & 0.03  & 0.02  & 0.03 \tabularnewline
\hline
\end{tabular}
\end{center}
\caption{Time profiling (in percent) of NBSymple in 3 of its versions (as labeled
by the capital letter) for the 4 main tasks, labeled as in Fig. 1, for the three values of $N$
given in the table heading.}
\label{tab4}
\end{table}


\begin{table}[t]
\small
\begin{center}
\begin{tabular}{l r  r  r}
\hline
\bf task &  $N_1$ &  $N_2$ &  $N_3$ \\
\hline
1&  94.0  &  99.03  &  99.9810 \\
2&  5.3    &  0.90    &  0.0165 \\
3&  0.5    &  0.06    &0.0018     \\
4&  0.2     &  0.02    &  0.0007 \\
\hline
\end{tabular}
\caption{Profiling for NBSympleC in its leapfrog, double precision, mode.
The entries are the time fractions (in percentage) spent by the code in performing the 4 main tasks (numbered 
as in Fig. 1 and Tab. 4) for three values of the number of particles: $N_1=480$, $N_2=30,720$, and $N_3= 1,536,000$.}
\label{tab5}
\end{center}
\end{table}

\begin{table}[t]
\small
\begin{center}
\begin{tabular}{l r  r  r}
\hline
\bf task &  $N_1$ &  $N_2$ &  $N_3$ \\
\hline
1 & 85.5  & 99.14  & 99.9819\\
2 & 13.1  & 0.80& 0.0163 \\
3 & 0.9 & 0.04  & 0.0012  \\
4 & 0.5  & 0.02    & 0.0006  \\
\hline
\end{tabular}
\caption{As in Tab. 5, for NBSympleC in its sixth order, double precision, mode.}
\label{tab6}
\end{center}
\end{table}

To have a reliable evaluation of the computing times, we ran the codes up to
$0.5T$ using a (constant) time step $\Delta t = 5\times 10^{-4}T$, this
means the code was run for $1000$ cycles.
This way, we have a robust average value of the time spent per cycle by the
various code versions in each task by the simple ratio of the total time to the number ($1000$)
of cycles.\\

Fig. 2 shows the (average) time required by the 
various versions of
\par\noindent 
NBSymple for a single time step integration with
the leapfrog method. For a system of $480\leq N \leq 1,536,000$ particles
(the largest value of $N$ used is quite representative of
the number of stars in a real, popolous, globular cluster).
In the case of the fully serial code on a single PE (NBSympleA) and of the OpenMP parallel code on the double quadcore host (NBSympleB), the test is limited to
a maximum value of N=7680 because larger values require too long CPU times.
Their computing times show clearly the expected $N^2$ behaviour.
The ratio (version A to version B) of their computational times indicates a speed-up about $\sim 3.6$ 
(the best obtainable being 8, as the number of PEs).\\
The most immediate sketch of the advantage in using the GPU is by comparing
the curve with empty squares to that with crosses, this latter referring to
NBSympleC, that is the version of the code where the $O(N^2)$ part of the
code is performed by a single TESLA C1060 GPU. The speed-up given by the GPU
ranges between $\sim 15$ (for $N=480$) and $\sim 330$ (for $N=7680$).
The NBSympleE code which exploits two TESLA GPUs becomes two times faster
than when using a single GPU, showing an almost perfect efficiency when
$N$ is large enough to overcome the overhead, as evident in Fig. 2 for
$N \lap 10^4$. The fastest version available (the one which uses two TESLA GPUs) 
requires one minute to accomplish a full integration time step for $N=1,536,000$; taking
into account the need of at least $2\times 10^3$ cycles in a crossing
time to have an acceptable precision, this means about 33 hours ($1.4$ days)
to simulate by direct summation the evolution over one internal crossing time 
of a globular cluster moving in the galactic potential, and about 211 hours 
(8.8 days) to follow a complete revolution of the system around the galactic center, 
if the cluster is moving quasi-circularly at the Sun distance from the centre.\\
Fig. 3 reports compared performances of the versions of NBSymple in
double-precision mode. The basic, single PE, version decays in performances of
a factor $\sim 1.5$ respect to the SP mode.
The ratio between time spent by the single PE version and the OpenMP version
remains similar to the SP case in Fig. 2.
Of course, the best performances are achieved by the full GPU NBSympleE code,
although the performance degradation respect to the SP version
is of a factor $\sim 18$, as expected
due to the characteristics of the TESLA C1060 architecture.\\
The symplectic 6$^{th}$ order, double precision performances are shown in Fig.
4. We decided to investigate the 6$^{th}$ order method performances
in its DP mode, only, because the use of high order symplectic
integration is done to keep high precision, which would be lost if using
SP arithmetics. The gain in precision is, of course, paid by a
significant loss in speed. At this regard, a useful indication comes from 
comparing Fig. 3 and Fig. 4  with Fig. 1. Considering the, fastest, NBsympleE version (the one 
exploiting 2 GPUs) we see that the ratios between computing times per cycle,
in the most significant range covered by all the figures ($4 \leq$ Log$N \leq 4.5$) 
scales by 2 factors of 10 for Log$N=4$ (the time required for a time cycle by the $6^{th}$ 
order DP code is about 10 times longer than by the $2^{nd}$ order DP code, which, 
in its turn, is about 10 times longer than the $2^{nd}$ order SP code) and by a 
factor of 20 and a factor of 12 for Log$N=4.5$.
The performance degrading when using the DP instead of SP with the $2^{nd}$ order 
leap-frog integrator is mainly explained  by the specific architecture of the TESLA C1060 GPU 
(1 DP processor per 8 SP processors), while the further degrading when using the $6^{th}$ 
order integrator is explained by both the 8 force evaluations per cycle needed and the 
repeated transfers from the host to the GPU device memory and back (a total of 16 per particle).

It is relevant noting, however, that the need of a high order symplectic code
is mostly for celestial mechanics applications (small values of $N$) where the
CPU time is dedicated to follow a long integration time (thousands of
orbital periods) rather than to the force computation which is 
(relatively) unexpensive because $N$ is small.
\subsection{Hardware and software double precision arithmetics}
The study of $N$-body system evolution is, in many cases, characterized by
the requirement of high precision in the computations. This for two main
reasons, that are different for small and large values of $N$.

For small values of $N$ (celestial mechanics) the high precision in the
acceleration calculation is required to avoid secular growth of the
error over the huge time extension of the integration; for large values
of $N$, when time integration is not too extended, precision is mainly
required to reduce the so called problem of cancellation of terms, i.e.
the error due to difference between numbers that are very similar. This
happens when evaluating the distances of bodies in the system in an inertial
reference frame, usually centered at the center of symmetry of the external potential
where the system is orbiting, in the case when $R<<d$, where $R$
and $d$ are the sizes of the system and its distance to the origin of the
reference frame, respectively.
The correct approach to the problem of error reduction is via coupling a
symplectic, high order, integration scheme to double or quadruple (in the case
of celestial mechanics applications) precision arithmetics.
At present, the NVIDIA TESLA C1060 GPU supports double-precision
64bit floating point arithmetics in a limited way, because
each of the processing units contains one DP processor alongside
the 8 SP processors. This means that there are only 30 DP units
available if doing DP only calculations.
The relevant performance degradation convinced us to seek for a possible solution to join
precision and speed. A straightforward way to emulate DP with single precision
artihmetics is done transferring the DP representation of every particle space coordinate 
to the GPU memory where it is shared into two SP allocations, one where the most significant digits
are stored and and another for the lesser significant digits. When forces have to be computed by the GPU,
it joins the two SP memory allocations to give a good DP emulation 
(14 digits instead of 16) yielding to a satisfactory round-off
error in the, delicate, evaluation of coordinate difference. Such a representation of DP is known as \lq double-single\rq~ precision
(DSP\footnote{http://crd.lbl.gov/~dhbailey/mpdist/}).\\
The loss in performance has been checked to be acceptable, and the implementation 
is not dependent on the particular hardware used.
The practical implementation of DSP in our code was done by mean of the SAPPORO library
(Gaburov et al. 2009).\\
Fig. 5 shows a performance comparison between the DP and DSP codes 
using the leapfrog integrator. It shows the (average) time spent for integrating the system for a single time step, as function of the number of bodies. It is evident that the DSP 
precision allows to integrate a system much faster than the DP precision.\\
\noindent
Finally, Tab. 7 shows the total energy conservation for the leapfrog and $6^{th}$ order methods after 1000 time steps.\\

\begin{table}[htdp]
\begin{center}
\begin{tabular}{|c|cc|cc|}
\cline{2-5} 
\multicolumn{1}{c|}{} & \multicolumn{2}{c|}{Leapfrog} & \multicolumn{2}{c|}{Sixth Order}\\
\hline
N & $|\Delta E/E_0|_{DP}$ & $|\Delta E/E_0|_{DSP}$ & $|\Delta E/E_0|_{DP}$ & $|\Delta E/E_0|_{DSP}$\\
\hline
480  & 3.39$\times10^ {-13}$ &  6.68$\times10^ {-10}$	& 3.09$\times 10^{-15} $ & 1.15$\times 10^{-10} $\\
960  & 1.57$\times10^ {-10}$ &  5.93$\times 10^{-11}$ & 0 & 5.10$\times 10^{-11}$\\
1920  & 6.75$\times10^ {-13}$ &  1.90$\times 10^{-11}$ & 3.04$\times 10^{-15} $ &  2.11$\times 10^{-11}$ \\
3840  &  8.60$\times 10^{-13}$ &  1.18$\times 10^{-11}$ & 8.10$\times10^ {-10}$ & 1.19$\times 10^{-11}$ \\
7680  &  1.12$\times 10^{-12}$ &  7.32$\times 10^{-12}$ & 1.01$\times 10^{-15} $ & 4.64$\times 10^{-12}$\\
15360  & 1.18$\times10^ {-12} $ & 2.73$\times 10^{-13} $ & 2.83$\times 10^{-15} $ & 5.12$\times 10^{-12}$\\
\hline
\end{tabular}
\caption{Relative errors in energy evaluated over 1000 time steps as $\Delta E/E_0=[E(1000\Delta t)-E(0)]/E(0)$.}
\label{tab7}
\end{center}
\end{table}

\subsection{Code time profiling}
In order to see whether and where it is possible to work to improve code performances, we
checked the time spent by the versions of the codes that actually run on GPUs (versions C, D and E)
in their different main tasks, which are essentially defined by the
flow-chart in Fig. 1 and numbered from 1 to 4. The profiling is resumed in Tabs. 4, 5 and 6.
As expected, the pairwise force calculation (task 1) requires a longer time at increasing $N$, rising to more than $99.8 \%$ for the largest $N$. Only the version C of the code (the one which gives to a single CPU the O($N$) computational load) spends a relatively significant fraction of time in other tasks than 1.
Due to that version E performs both pairwise and external force calculations on the GPU, no distinction is possible between tasks 1 and 2, and this explains the absence of values in the task 2 row for this code version.

\section{Some simulation tests}

\subsection{Quasi circular cluster orbits in the Milky Way}

In Figs. 6, 7 and 8 we display some snapshots of the projections onto the $x-y$ coordinate plane of a cluster composed by $N=15,360$  equal mass stars moving very close to the Milky Way plane of simmetry on a quasi circular orbit with radius $\sim 8$ kpc, simulated with NBSymplE using two TESLA C1060 GPUs. Fig. 6 shows a clear, quick development of a tidal tail in less than one orbital period
(here, one orbital period is about $60T$, where $T$ is the internal crossing time, we adopted as time unit).
Fig. 7 is a zoom of the cluster configuration at about half revolution around the galactic centre: note the evident barlike structure in the inner zone. After about 1.5 revolutions around the galactic centre, the cluster shows two extended tails along its orbit, with two clumps, one in the leading and one in the trailing tail. These clumps have already been noted in previous high precision $N$-body simulations of globular clusters moving in an external field and their explanation is found in Capuzzo-Dolcetta et al. (2005)
and Di Matteo et al. (2005) as due to a local deceleration of the stellar motion, causing an effect similar to a ``motorway traffic jam" (essentially, if we compare the stellar motion to a fluid stream, for which the continuity equation holds, along the tail an overdensity occurs whenever $\nabla \cdot {\bf v}<0$, where ${\bf v}$ is the colective stellar velocity). It is relevant noting that similar clumps have actually actually been observed in real globular clusters, the most know example being Pal 5 (see Odenkirchen et al. 2003). 

\subsection{Quasi radial cluster-massive black hole collisions}

For the sake of testing of the capabilities of our code we ran some ``stiff" simulations, i.e. those of the face-on collision of a star cluster with a massive black hole sited at the centre of the galaxy, during its quasi radial motion in the central part of the Galaxy. 
These simulations, at varying the mass ratio between the black hole (represented as a further massive point) and the star cluster and at varying the number of stars in the cluster have a great astrophysical interest because they deal with the process of strong interaction with a massive central object that orbitally decayed globular clusters may have actually suffered  in their motion in the parent galaxy. Apart from this scientific relevance, these simulations constitute a serious and difficult test for an $N$ body code to face. Actually, the close interaction of a star system composed by $N$ point masses with a single body (the ``black hole") much heavier than the individual cluster objects is hard to be followed by a numerical code because the very large acceleration induced by the massive black hole on passing-by cluster stars may cause an exceedingly large energy error. The energy error can, partly, be controlled by a proper reduction of the time step which cannot be, however, reduced below a threshold under which cumulation round-off error and CPU time get exceedingly large. The difficulty of such simulation grows at reducing the ``collision" impact parameter $b$, i.e. the minimum distance between the cluster barycenter and the black hole during the cluster oscillatory motion around the massive object. 
To maximize the sharpness of collision we chose almost radial trajectories ($b\simeq 0$) for the cluster moving in the inner part of the Galaxy. A summary of results of our simulations at varying the black hole to cluster mass ratio  is given in Table 8.
Fig. 9 shows the motion of the cluster on the $x,y$ plane at various times, showing the very fast development of the cluster ``arc"-like shape around the apocenter.

\begin{table}[!h]
\small
\begin{center}
\begin{tabular}{cll}
\hline
$M_{bh}/M$ & ${\Delta E}/{E_0}$ & ${\Delta L}/{L_0}$\\
\hline
0.1& 3.06$\times10^{-7}$ & 5.90$\times10^{-4}$\\
1& 7.75$\times10^{-6}$& 1.06$\times10^{-2}$\\
50& 9.23$\times10^{-5}$& 2.65\\
105& 5.63$\times10^{-3}$& 3.91\\
\hline
\end{tabular}
\caption{Relative variations in energy and angular momentum over the time interval [0,4], which corresponds to about 6.4 complete radial oscillation through the galactic center of an $N=15,360$ star cluster. The ratio between the black hole mass and the cluster mass is $M_{bh}/M$. The motion starts with cluster barycenter at $(x_0,y_0,z_0)=(10,0,0)$ with initial velocity $(\dot x_0,\dot y_0,\dot z_0)=(0,0.001v_c,0)$, where 
$v_c$ is the local circular velocity. The number of time steps is $4\times 10^4$; the total CPU time $\simeq 9.1\times 10^4$ sec.}
\label{tab8}
\end{center}
\end{table}

\section{CODE AVAILABILITY}
We make available to the community the basic version of our code.  The web site where the code, as well as instructions for running it plus some other useful information, are found is:\par\noindent
astrowww.phys.uniroma1.it/dolcetta/nbsymple.html

\begin{thebibliography}{00}


\bibitem{} Allen, C., Santillan, A., 1991, RMxAA, 22, 255
\bibitem{} Barsdell, B. R., Barnes, D. G., Fluke, C. J., 2010, Advanced Architectures for Astrophysical Supercomputing, eprint arXiv:1001.2048
\bibitem{} Belleman, R. G., B{\'e}dorf, J., Portegies Zwart, S. F., 2008, High performance direct gravitational N-body simulations on graphics processing units II: An implementation in CUDA, New Astronomy, Vol. 13, pp103-112
\bibitem{} Capuzzo--Dolcetta, R., Di Matteo, P., Miocchi, P. 2005, AJ, 129, 1906 
\bibitem{} Capuzzo--Dolcetta, R., Pucello, N., Rosato, V., Saraceni, F.,
2001, JCP, 174, 208
\bibitem{} Di Matteo, P., Capuzzo--Dolcetta, R., Miocchi, P. 2005, Cel. Mech., 91, 59
\bibitem{} Elsen, E., Houston, M., Vishal, V., Darve, E., Hanrahan, P., and Pande, V.S.
2006, N-Body simulation on GPUs. Proceedings of the 2006 ACM/IEEE conference on Supercomputing, 188 
\bibitem{} Gaburov, E., Harfst, S., Portegies Zwart, S., 2009, NewA, 7, 630-637
\bibitem{} Hamada, T.,  \& Iitaka, T. ,  2007, The Chamomile Scheme: An Optimized Algorithm for N-body simulations on Programmable Graphics Processing Units, eprint arXiv:astro-ph/0703100
\bibitem{} Kinoshita, H., Yoshida, H., Nakai, H., 1991, CeMDA, 50, 59
\bibitem{} Nyland, L., Harris, M., Prins, J., 2007, \emph{ ``Fast N-Body
Simulation with CUDA"}, GPU Gems 3, H Nguyen, ed., Prentice-Hall 2007
\bibitem{} Odenkirchen, M., et al. 2003, AJ, 126, 2385 
\bibitem{} Plummer, H.C., 1991, MNRAS, 71, 460
\bibitem{} Portegies Zwart, S.~F., Belleman, R.~G., Geldof, P.~M., 2007, NewA,
12, 641
\bibitem{} Yoshida, H., 1991, Proceedings of the 24th Symposium on
Celestial Mechanics, held in Tokyo, H. Kinoshita and H. Yoshida eds.
(Tokyo, Japan) 132

\end{thebibliography}

\end{document}